\documentclass[12pt,a4paper]{article}

\usepackage{amsmath,amssymb,graphicx,epsfig}

\usepackage{color}

\begin{document}

\title{\bf The Demon in a vacuum tube}

\author{Germano D'Abramo\vspace{0.2cm}\\
{\small Istituto Nazionale di Astrofisica,}\\ 
{\small Via Fosso del Cavaliere 100,}\\
{\small 00133, Roma, Italy.}\\
{\small E--mail: {\tt Germano.Dabramo@iaps.inaf.it}}}
\vspace{0.2cm}
\date{}

\maketitle

\begin{abstract}
In the present paper, several issues concerning the second law of thermodynamics, Maxwell's demon and Landauer's principle are dealt with. I argue that if the demon and the system on which it operates without dissipation of external energy are made of atoms and molecules (gas, liquid or solid) in thermal equilibrium (whose behaviour is described by a canonical distribution), then the unavoidable reason why the demon cannot successfully operate resides in the ubiquity of thermal fluctuations and friction. Landauer's principle appears to be unnecessary. I also suggest that if the behaviour of the demon and the system on which it acts is not always describable by a canonical distribution, as would happen for instance with the ballistic motion of electrons at early stages of thermionic emission, then a successful working demon cannot be ruled out a priori. A critical review of two recent experiments on thermionic emission Maxwell's demons is also given.\\


\end{abstract}

\section{Introduction}

Classical thermodynamics was essentially born in the nineteenth century and its original goal was to understand the fundamental principles that underlie the operation of heat engines with the aim of making them as effective as possible (note that the 1824 seminal work by Sadi Carnot, who is considered as the father of thermodynamics, deals with heat, power energy and engine efficiency).

Heat engines are machines having almost always a gas as working substance and performing cyclic transformations on the Pressure--Volume ($PV$) diagram. Hence, the main subject of classical thermodynamics is the heat and work exchange between the environment and systems whose state is uniquely defined by Volume, Pressure and Temperature, together with the equation of the state of the working substance.

One cannot deny that early classical thermodynamics mainly dealt with a gas inside a cylinder, performing $V,P,T$ transformations. In such a context, the second law of thermodynamics (Kelvin--Planck formulation) appears to be even a straight consequence. Let me explain. We all know that for a transformation to be practically exploitable as a source of work, it must be cyclical. Moreover, the work performed by a gas during a cycle is equal to the area inside the closed loop on the $PV$ diagram that represents the cyclic process---the work is positive, \emph{i.e}.,~released, if the cycle is clockwise. (Actually, this is true for all working substances, but only gases can have significant variations in both $P$ and $V$.) Thus, the only possibility for a cyclic process to perform work is to be a closed loop enclosing a non-zero area on the $PV$ diagram. This is possible only if the system follows more than one isothermal on the $PV$ diagram for a non-zero portion in the way back to its initial state and thus only if it exchanges heat with heat reservoirs at more than one absolute temperature. Note that a loop cannot be closed using only adiabatics (transformations with no heat exchange) since they do not intersect each other. Obviously, one could have a more complex machine working with more than a single cylinder. In this case the machine cannot be represented on a single $PV$ diagram and the above approach cannot be trivially applied (it could be interesting to investigate whether the above approach could be extended to such engines).

Thus, in classical thermodynamics the second law appears to be a strict necessity, even a trivial one. As a personal experience at high school and university, I never saw a derivation of the second law like that given here. In almost all textbooks, the second law was introduced as a ``deus ex machina'', apart from the Clausius formulation that is actually more intuitive and linked to everyday experience.

Serious problems with the second law began with the modern development of the atomic theory of matter and the birth of statistical mechanics around 1870. When the theory of statistical physics was developed by Maxwell, Boltzmann and others, it became clear that the second law of thermodynamics could not hold unconditionally, but only statistically. The Brownian motion is a well-known macroscopic example of that, as early noted by Poincar\`e. In other words, entropy of isolated systems is not forbidden to decrease, but in all processes the {\em probability of continuous and macroscopically significant (and also able to provide usable work) entropy decrease is extremely small}.

The infamous Maxwell's demon first appeared in that period. Maxwell's thought experiment shows a clever way to macroscopically amplify such a statistical (microscopic) breach to the second law. Since then, several attempts have been put forward to exorcise the demon and save the second law. These attempts span fluctuations (Smoluchowski), measurement entropy generation (Szilard, Brillouin) and information theory (information erasure entropy, Landauer and Bennett). A detailed review of and a really thorough reference list on Maxwell's demon and various exorcisms can be found in~\cite{ler,io4}.

Nowadays, Laundauer's principle seems to be widely acknowledged as the ultimate and definitive exorcism of Maxwell's demon. The demon, every conceivable demon, in order to perform a cyclic process needs to erase information acquired (and stored) on previous phases of the cycle. According to the information entropy orthodoxy, this information erasure always generates a thermodynamic entropy increase greater than or equal to the alleged entropy decrease operated by the demon. Personally, I confess to have always been uncomfortable with this explanation. I ever failed to see the fundamental, ultimate physical connection between information entropy and thermodynamic entropy, with the exception of the mathematical formalism (the Shannon entropy has the same mathematical form of the Boltzmann entropy). However, close parallels in mathematical formalism do not always necessarily mean ``same underlying physics'', or that one theory is the fundamental basis of the other. For instance, electromagnetic waves and acoustic waves are described with nearly the same mathematics, but electromagnetic~waves are neither acoustic waves nor the fundamental basis of the acoustic waves (or vice versa). Note that I am not denying that information is physical and that information erasure may imply entropy increase. I am saying that this cannot be the most basic reason for the failure of every~demon.

In the following section, I shall describe a modified Szilard's engine (already described in~\cite{io4}) and show how Landauer's information-theoretic principle seems to fall short of the initial expectations of being the only fundamental ``physical'' reason of all Maxwell's demons failure. The modified Szilard engine operates without any steps in the process resembling the creation or destruction of information. I argue that the information-based exorcisms must be wrong, or at the very least superfluous, and that the real physical reason why such engines cannot work lies in the ubiquity of thermal fluctuations (and friction). {Other Szilard-like engines not requiring measurements or memory erasure have already been described in the literature, see for instance \cite{ber}.}

%%%%%%%%%%%%%%%%%%%%%%%%%%%%%%%%%%%%%%%%%%%%%%%%%%%%%%%%%%%%

\section{The Exorcism Is in Fluctuations/Friction}

\begin{figure}[t]
\centerline{\includegraphics[width=5cm]{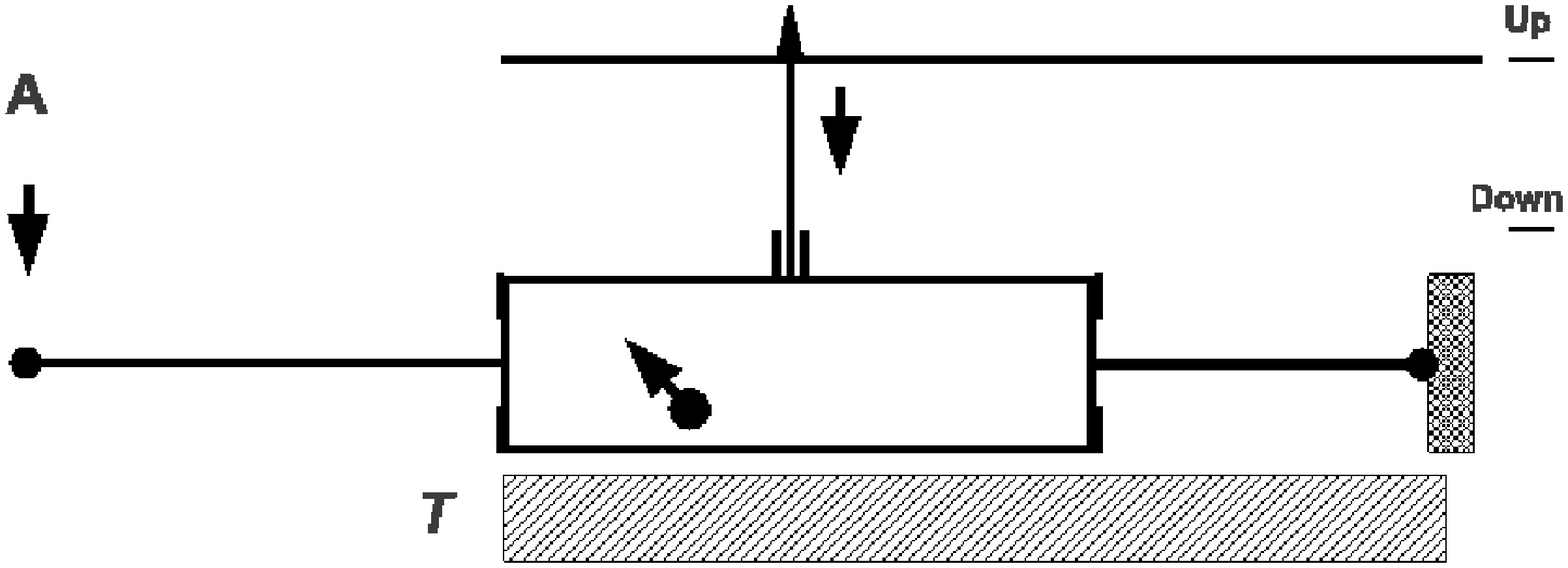}}
\vspace{0.5cm}
\centerline{\includegraphics[width=5cm]{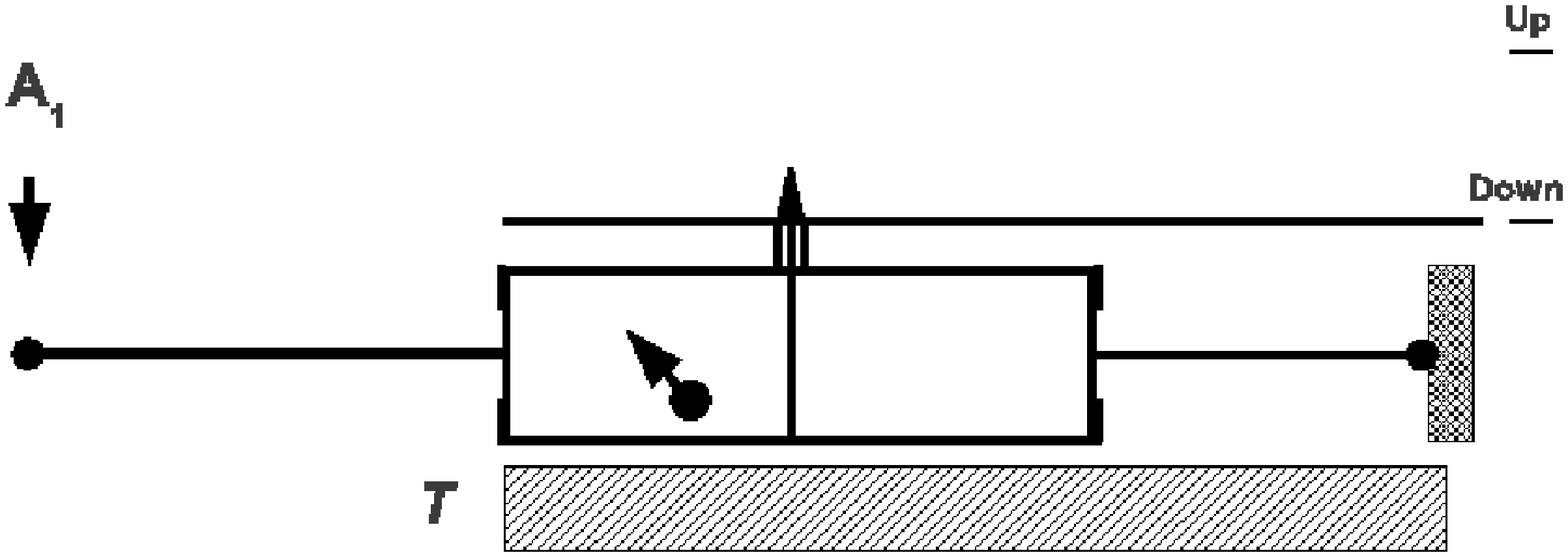}}
\vspace{0.5cm}
\centerline{\includegraphics[width=5cm]{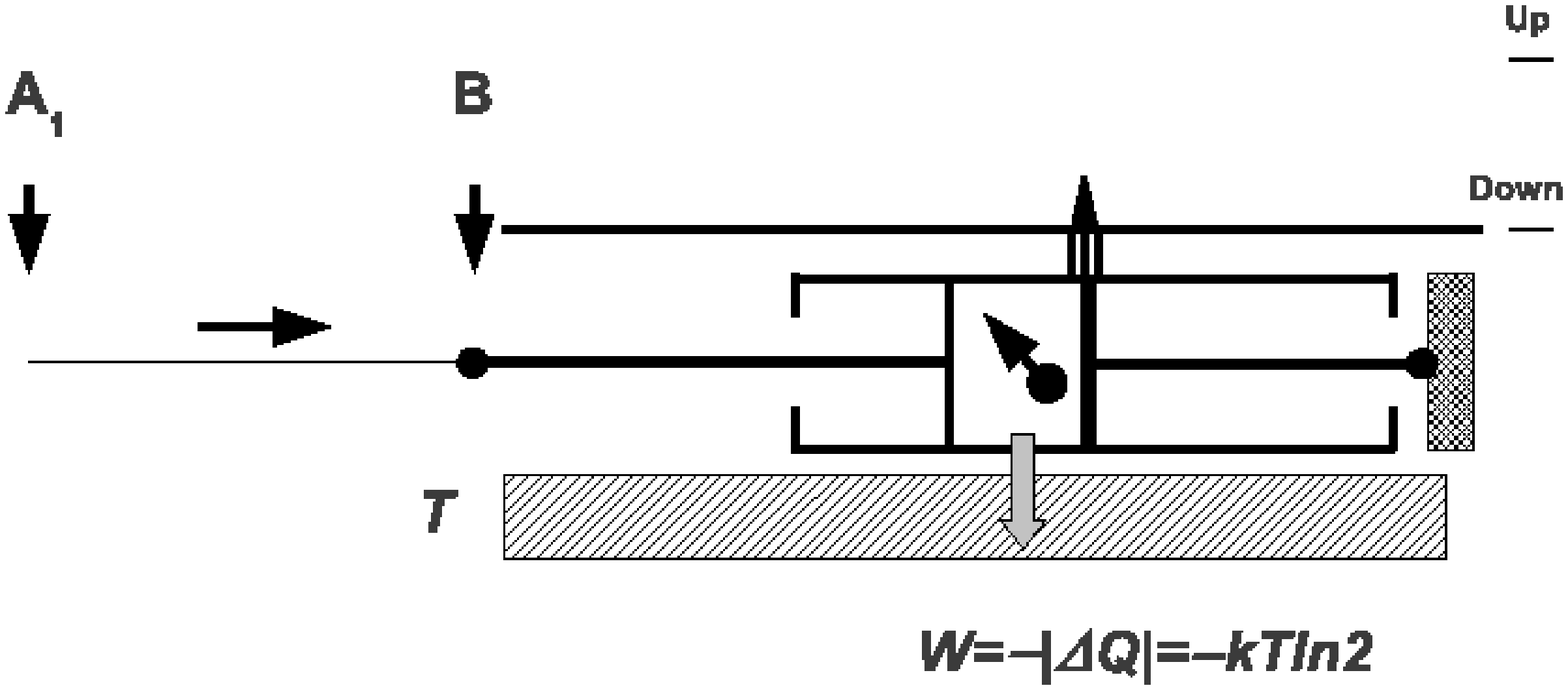}}
\vspace{0.5cm}
\centerline{\includegraphics[width=5cm]{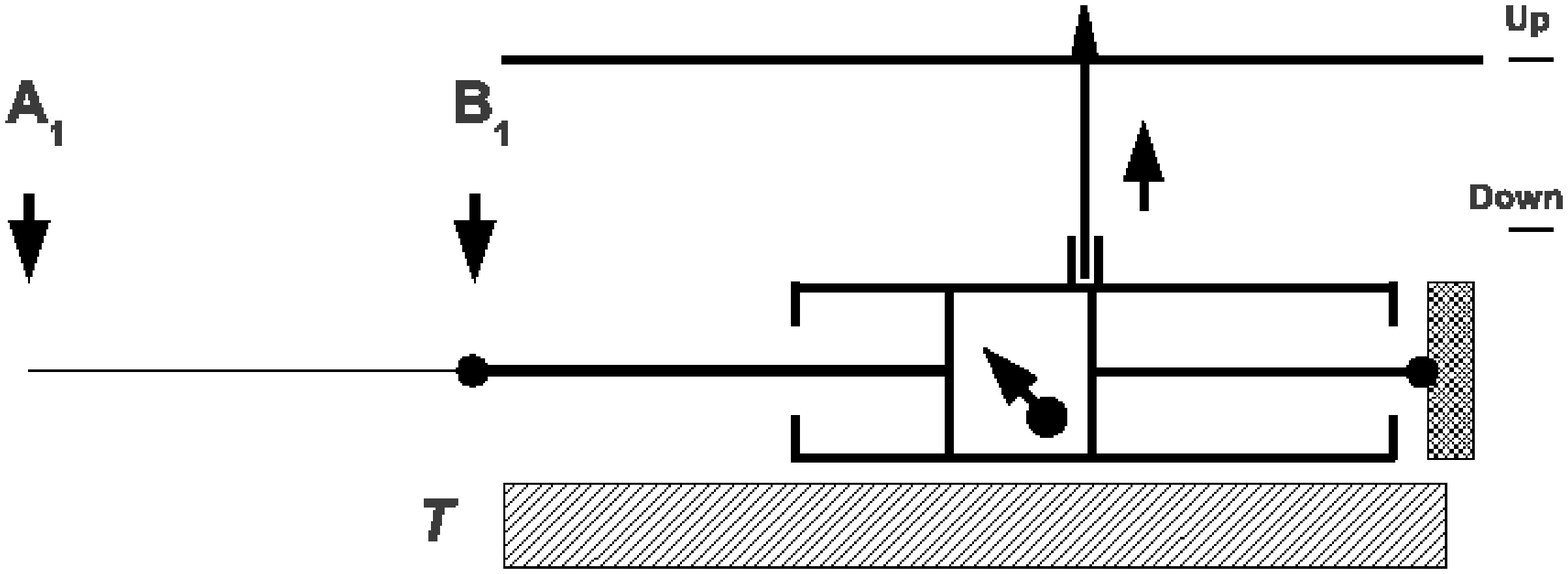}}
\vspace{0.5cm}
\centerline{\includegraphics[width=5cm]{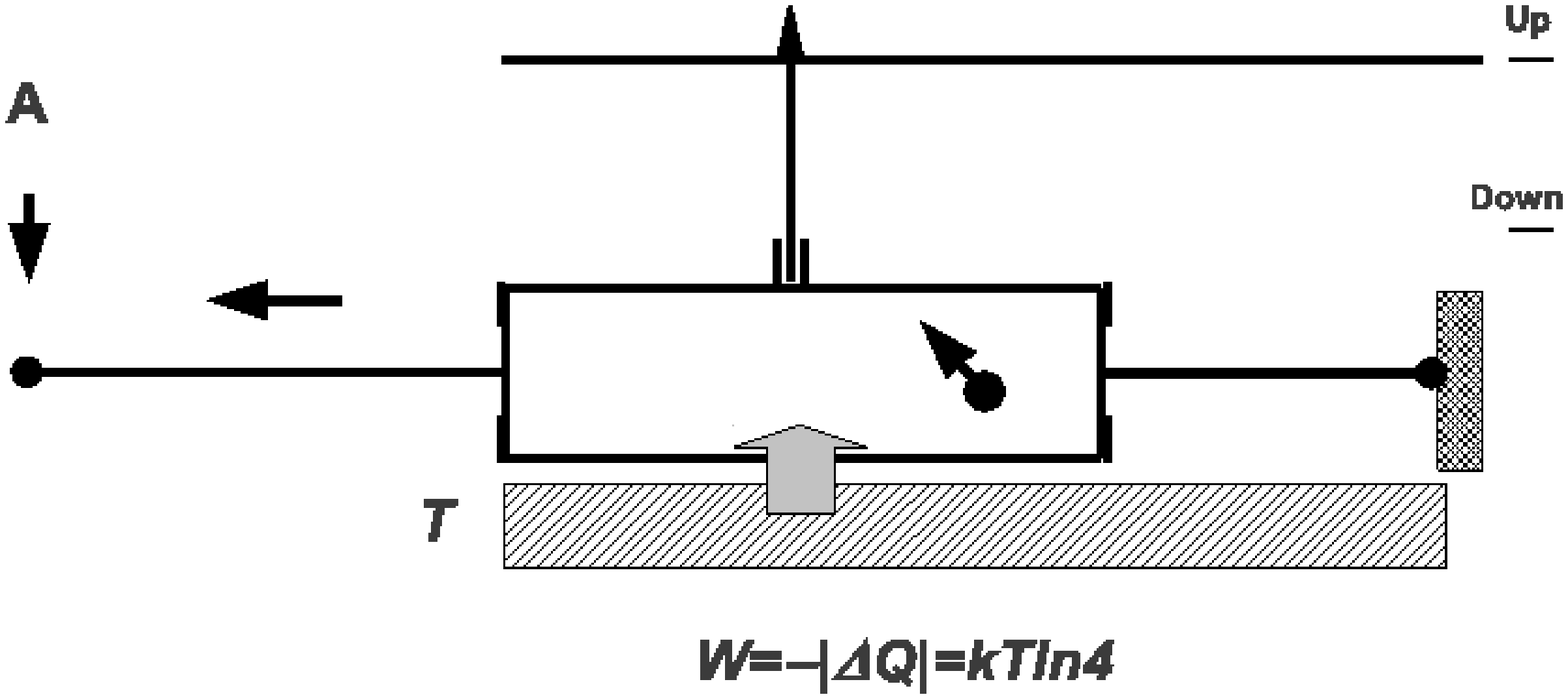}}
\caption{The modified Szilard's heat engine described in the text.}
\label{fig0}
\end{figure}

It is not difficult to devise a modified Szilard's heat engine that cyclically works without the need of information acquisition and/or memory erasure. Such an engine is shown in Figure~\ref{fig0}. It is made of a movable cylinder and two pistons (the left one movable and the right one fixed). There is also a partition that can be lowered in the middle of the cylinder and that can slide horizontally on a lowering rod without friction as the cylinder moves. The insertion of the partition involves no work or heat. All the mechanical parts are thought to be without friction, as has been done extensively in the literature on the subject (more on this later). Initially the entire volume $V$ of a cylinder is available to a single molecule (step A). The behaviour of the molecule is described by the equation of state $PV=kT$. Then, the partition is lowered into the cylinder, dividing it into two equal chambers (step A$_1$). The molecule is trapped in one of these two chambers.

Then, the movable piston is pushed infinitesimally slowly (reversibly) to position B and the one-molecule gas undergoes an isothermal compression from $V/2$ to $V/4$. The work $W_{A_1\to B}$ externally done to the gas is equal to $kT\ln 2$, which is also equal to the heat transferred by the gas to the heat reservoir at temperature $T$. Note that this part of the cycle is independent of which side of the partition the molecule is on at step A$_1$, hence we do not need any information acquisition (with subsequent memory erasure). The final position of the movable piston is always at point B, no matter if the molecule is in the right or in the left chamber after step A$_1$. Besides, the movement of the partition can be mechanically coupled to the cyclic movement of the movable piston, and thus without any need of information acquisition and/or memory erasure to operate the partition itself.

One may complain that the compression procedure depends on whether the molecule is trapped on the left or right. Namely, if the molecule is on the left, the piston moves the whole cylinder first, with its action on the cylinder mediated by the gas pressure. If the molecule is on the right, the piston moves in unimpeded to contact the partition and then compresses the gas. Since the two processes appear to be slightly different, one may wonder that in order to operate the device, one has to know which condition is at hand. This would mean measurement and/or memory erasure. Under a more careful analysis one can easily see that the two processes are not different at all. In both cases there is a first phase where the device moves unimpeded until the right piston, if the molecule is on the left, or the left piston, if the molecule is on the right, touches the partition (from step A$_1$ to the midpoint between A$_1$ and B), and then a second phase in which there is the true gas compression (from the midpoint between A$_1$ and B to step B). These two phases are physically perceived always in the same way by who/what operates the device: the first half of the compression is always equally ``loose'', no matter where the molecule is at the beginning of the process, while the second half is the true gas compression. At step B$_1$, the partition is raised and the cycle is completed with an isothermal expansion from $V/4$ to $V$ (with movable piston again in position A). Now, the work $W_{B_1\to A}$ made by the gas to the environment is equal to $kT\ln 4$, which is also equal to the heat transferred by the heat reservoir at temperature $T$ to the gas. The net work output $W_n$ over any cycle is then equal to $W_{B_1\to A}-W_{A_1\to B}=kT(\ln 4 -\ln 2)=kT\ln 2$. Moreover, the entropy variation $\Delta S$ of the entire system (engine + reservoir) is equal to $-k\ln 2$.

If we want to save the second law in the above scheme, then some other mechanisms must come into play to prevent the modified Szilard engine from operating. For instance, thermal fluctuations surely afflict the mechanical parts of the engine (pistons, partition and so on). The pistons must be sensitive to energy of the order of $kT$, the mean energy of the molecule, and they themselves are plagued by fluctuations of order $kT$, like the Smoluchowski one-way valve. Actually, if there were no friction, then the device could operate even with arbitrarily massive pistons, partition and cylinder (massive means not instantaneously sensitive to energy of the order of $kT$). As a matter of fact, without friction even the tiny kick of a single molecule can move a massive piston/cylinder (conservation of linear momentum). But friction cannot be eliminated, even ideally, since thermal fluctuations of the matter along the contact points between the pistons' edge and the cylinder's walls originate an unavoidable friction force that is surely greater than the force imparted by the molecule to the pistons.

However, if such effects afflict our modified Szilard's engine, then the same effects must afflict the original Szilard's engine, since both engines are mechanically similar. Hence, the appeal to information acquisition and/or memory erasure entropy costs to defeat the Maxwell's demon in the instantiation of the original Szilard's engine is superfluous. On the other hand, if information acquisition and/or memory erasure entropy costs are strictly necessary to defeat original Szilard's engine, then this means that no other mechanisms are able to prevent its operation. But this last thing would necessarily apply also to our modified Szilard's engine. Thus, our engine would surely violate the second law, since measurement and memory erasure, with their associated entropy costs, do not apply to it, as we saw above. As a logical consequence, measurement and memory erasure entropy costs are again unnecessary to defeat Maxwell's demon, this time in the instantiation of our modified Szilard's engine.

As a conclusion, the appeal to measurement and memory erasure entropy costs made by Szilard and Bennett within the original Szilard's engine appears to be an arbitrary choice rather than a necessity in defeating the Szilard's demon.

Probably, the true reason why the demon cannot operate, namely cannot macroscopically violate the second law and create usable work, is the ubiquity of thermal fluctuations and friction in the physical matter, the matter that inevitably constitutes both gas and every device conceived to sort molecules. If the demon and the system on which it operates without dissipation of external energy are made of atoms and molecules in thermal equilibrium (whose behaviour is described by a canonical distribution), then the fluctuations/friction exorcism is unavoidable. See also a more general principle introduced by J.~Norton in a recent paper~\cite{wai}.

%%%%%%%%%%%%%%%%%%%%%%%%%%%%%%%%%%%%%%%%%%%%%%%%%%%%%%%%%%%%

\section{The Demon may Lurk in a Vacuum Tube}

In the previous sections, I have shown that if one only deals with classical thermodynamics, \emph{i.e}.,~with a gas, a system status defined by macroscopic variables $V,P,T$ and transformations on a $PV$ diagram, then the second law in the Kelvin--Planck formulation is true by necessity. Problems begin to emerge when one looks at the second law from a statistical mechanics point of view. The second law loses its absolute value, it could not hold unconditionally, but only statistically (it is exactly this statistical breach that the demon puts its fingers on). In this last case, I have also argued that thermal fluctuations and friction, rather than information-theoretic principles, are the unavoidable reason why the demon cannot operate, at least if the demon and the system on which it acts without dissipation of external energy are made of matter whose behaviour is fully described by a canonical distribution---again, gas, liquid, solid in thermal equilibrium. Thermal equilibrium is essential for a genuine Maxwell's demon because if we have a thermal gradient, the demon's action may be ascribed to a possible extraction of work from heat reservoirs at different absolute temperatures (like a standard Carnot engine), which does not count as a ``regular'' second law violation.

Here I want to go further and suggest that if the demon and the system on which it acts are not always describable by a canonical distribution, then a successful working demon cannot be ruled out a priori.

In a series of recent papers~\cite{io4,io,io2, io2-1}, I have described in detail a vacuum spherical capacitor (a special kind of vacuum tube) that generates a macroscopic voltage between its spheres harnessing the heat from a single thermal reservoir at room temperature. {It is a kind of demon that sorts electrons rather than molecules.}
The working principle is trivial and it is based on the thermionic emission imbalance between two concentric spherical electrodes with different work functions. The inner one is coated with a well known photo-cathode, Ag--O--Cs, with very low work function. The outer one is made of a common metal with relatively high work function.

The unbalanced behaviour of the electrons {\em just after} the emission from the Ag--O--Cs coating is governed by the mechanical/ballistic laws of motion and not by the canonical distribution that describes systems in thermal equilibrium, $p(\mathbf{x},\mathbf{p})=\frac{e^{-E(\mathbf{x}, \mathbf{p})/kT}}{Z}$, where $E(\mathbf{x},\mathbf{p})$ is the energy of the system, $Z$ is the normalizing partition function, and the multi-component $\mathbf x$ and $\mathbf p$ are generalized configuration and momentum coordinates. Hence, the randomizing (disruptive) effect of thermal fluctuations for standard demons does not appear to apply here. Obviously, if the thermo-charged capacitor is left open-circuited (terminal leads not shunted by a load) a dynamical equilibrium of electrons (space charge inside the capacitor) is eventually reached and this final state should be describable by a suitable canonical distribution.

Moreover, the functioning of such a device is not representable on a $PV$ diagram, which otherwise would have implied the impossibility of second law violation, as explained in the first section.

In what follows, I describe two recent instantiations of such a vacuum tube demon.

\subsection{Recent Maxwell's Demon Instantiations}

In recent years, two interesting experiments have been put forward, and actually carried out, allegedly instantiating Maxwell's demon. In both experiments, use is made of thermionic emitted electrons bended by an external static magnetic field.

\begin{figure}[t]
\centering
\includegraphics[height=4cm]{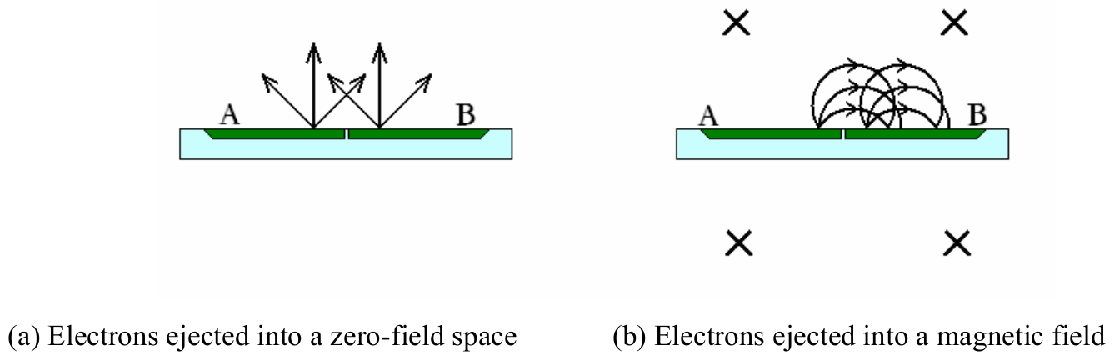}
\caption{Sketched side views of the two similar electrodes in Fu and Fu vacuum bulb~\cite{fu}.}
\label{fig1}
\end{figure}

The first experiment is by Fu and Fu~\cite{fu}. Their device is made of two similar (\emph{i.e}.,~bearing almost equal work functions, $\phi_A\simeq\phi_B$) plates covered with photo-cathode Ag--O--Cs, placed side by side inside a vacuum bulb at room temperature and electrically insulated from one another (see Figure~\ref{fig1}). The bulb is immersed in a static uniform magnetic field generated by a permanent magnet. During the experiment, the magnet is gradually moved toward the bulb (which is shielded from electromagnetic ~waves and external electric field by a copper box) from relatively high distance (${\bf B}\simeq{\bf 0}$). During the magnet move, the external magnetic field is non-static and thus some induced currents may arise. It is not clear from the paper whether the published measurements have been performed after a suitable relaxation time.

According to the authors, the external magnetic field plays the role of the demon, bending the otherwise symmetric electronic flux preferably toward one of the electrodes and hence generating a voltage drop (and a current when the bulb electrodes are shunted by a load). The authors claim to have measured a maximum current of the order of $10^{-14}$~A and maximum voltage drop of the order of $10^{-4}$~V.

The second experiment is by Perminov and Nikulov~\cite{pe}. It bears interesting similarities with that described above. Their device is sketched in Figure~\ref{fig3}. Perminov and Nikulov make use of two electrodes with explicitly different work functions, separated by a dielectric (insulating) surface. The apparatus is put inside a vacuum chamber at a uniform temperature of about $100 \,^{\circ}$C. Also in this case, there is an external static magnetic field. It is generated by a steady current flowing inside a linear conductor parallel to the dielectric surface. The magnetic field is supposed to act as a demon, pushing the thermionic emitted electrons from the low work function electrode to the high work function one and thus helping to overcome the electrodes' contact potential difference (see {Figure~\ref{fig3}).

\begin{figure}[t]
\centering
\includegraphics[height=6cm]{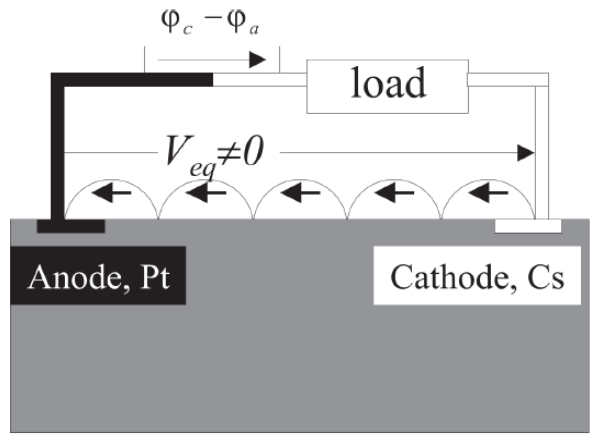}
\caption{Working principle of the Perminov and Nikulov device~\cite{pe}. The original figure caption in~\cite{pe} reads: ``The potential difference created because of the direct equilibrium movement of charged particles in magnetic field over the dielectric surface can exceed the contact potential difference between metals of the cathode and the anode, $V_{eq}> \phi_a -\phi_c$. Therefore a direct current in the electric circuit containing a load is possible under equilibrium conditions.''}
\label{fig3}
\end{figure}

As a matter of fact, in almost all textbooks~\cite{cu} it is said that when two materials with different work functions, $\phi_A$ and $\phi_B$ with $\phi_A>\phi_B$, are joined at one end, a voltage drop equal to $\Delta V =\frac{\phi_A-\phi_B}{e}$ builds up not only across the joined surfaces (contact potential) but also instantaneously between the free ends of the materials (see Figure~\ref{fig2}) where charges also accumulate, and it is opposite to the voltage built up at the contact surface. {Note that such a voltage drop is not intended in the textbooks to be that generated by thermionic emission in the vacuum gap. It allegedly forms before and instantaneously as a consequence of the physical junction at one end (J-I, Figure~\ref{fig2}; more on this later).}
This voltage drop would prevent any thermionic emitted electron in the gap from spontaneously moving from the low work function free electrode surface to the high work function one.
\begin{figure}[t]
\centering
\includegraphics[height=6cm]{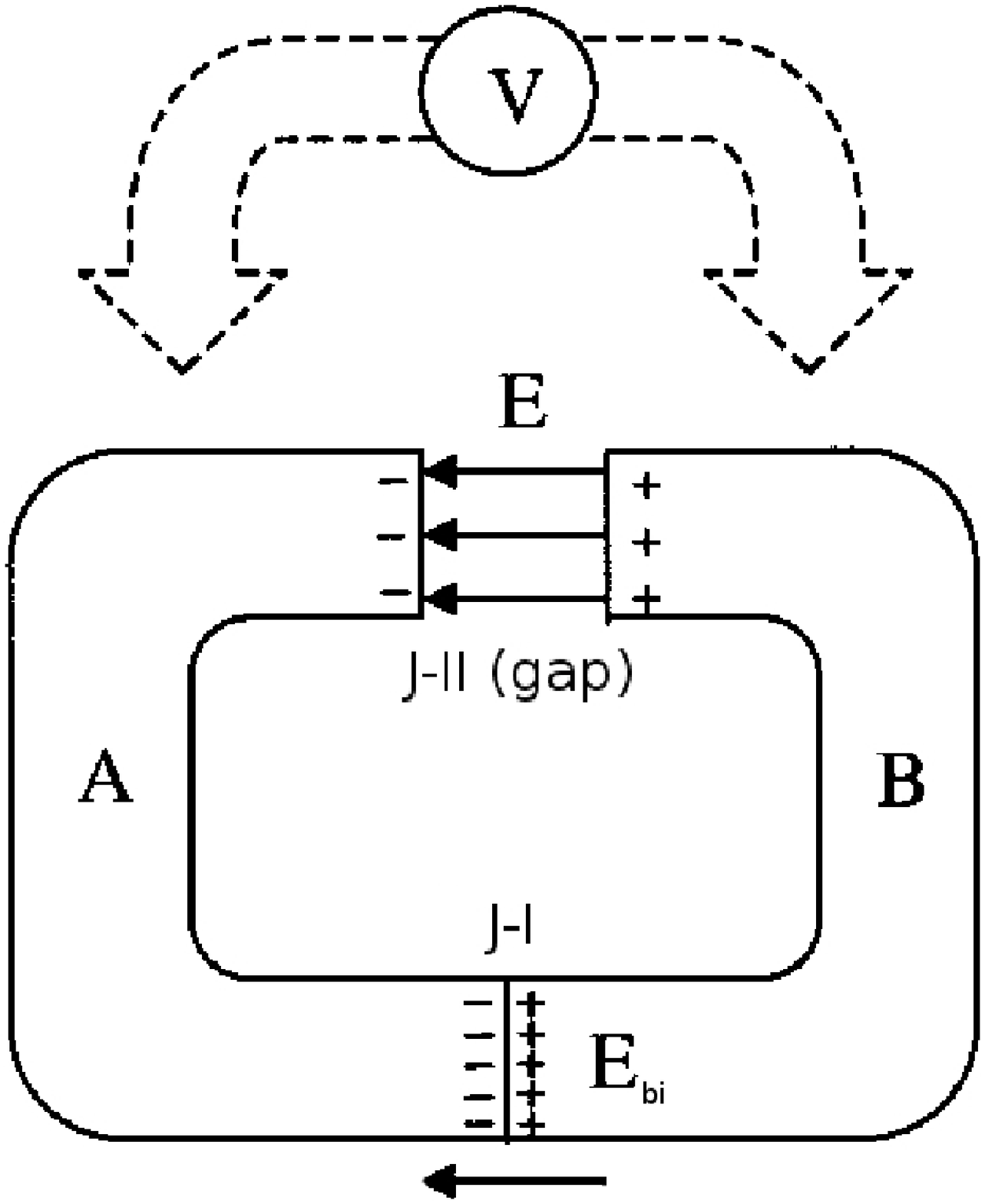}
\caption{Circuit of two connected materials A and B in vacuum. This scheme holds for metal to metal and metal to semiconductor junctions. Work functions are such that $\phi_A > \phi_B$. J-I is the physical junction while J-II is the gap (free ends). (Adapted from~\cite{cu}).}
\label{fig2}
\end{figure}
Even in the Fu and Fu experiment the two electrodes cannot obviously have $\phi_A = \phi_B$ (and thus one must assume $\phi_A < \phi_B$ or $\phi_A > \phi_B$) because the Ag--O--Cs coatings are manufactured directly inside the bulb with cesium deposition on a silver layer and oxidation. Such a process cannot obviously guarantee exactly $\phi_A =
\phi_B$. This means that even inside the Fu and Fu sample bulb a voltage drop $\Delta V =\frac{\phi_B-\phi_A}{e}$ (or $\Delta V =\frac{\phi_A-\phi_B}{e}$) builds up between the free surfaces of the electrodes and represents a potential barrier to the thermionic electrons. Thus, in both experiments it seems that the external static magnetic field helps pushing electrons across a potential barrier.

The actual experiment by Perminov and Nikulov has not been performed with the materials indicated in Figure~\ref{fig3}. The authors used two equal tungsten electrodes (thus, $\phi_A \simeq \phi_B$) and in any case they claim to have measured a current of the order of $10^{-7}$ A (across a circuit resistance of 1 M$\Omega$). They also assert that such a current is not ascribable to Thomson or Seebeck effects.

\subsubsection{Discussion}

I think that there is at least one weak point undermining both experiments, if not on the experimental/measurement side, at least from the theoretical modelling point of view. The point is that a static uniform magnetic field cannot perform work. The Lorentz force $\textrm{\bf F}= -e\,\textrm{\bf v}\times\textrm{\bf B}$ ($-e$ is the electronic charge), by definition, cannot perform work and thus it cannot be responsible for the overcoming of the mutual potential difference (contact generated) between the free electrodes in both experiments (see last paragraphs of previous section). As a matter of fact,

\begin{equation}
dW=d{\bf x}\cdot{\bf F}=-e\, dt\,{\bf v}\cdot ({\bf v}\times{\bf B})=0
\label{eq1}
\end{equation}
since for every vectors ${\bf a}$ and ${\bf c}$, we always have ${\bf a} \cdot ({\bf a}\times{\bf c})=0$.

Even if it were exactly $\phi_A = \phi_B$ (like in the Fu and Fu experiment as they claim, but this also holds for the Perminov and Nikulov device on which tests were actually carried out), the static magnetic field would not be able to generate a voltage drop and a current between the terminals of the device. In the Fu and Fu experiment, for instance, the bulb is a closed space where an equilibrium space charge eventually settles. Without magnetic field, electrons are emitted and absorbed by both Ag--O--Cs coatings in a sort of steady state equilibrium (space charge is just that). The glass envelope of the bulb contributes to electron confinement (see Figure~\ref{fig4}).

\begin{figure}[t]
\centering
\includegraphics[height=6cm]{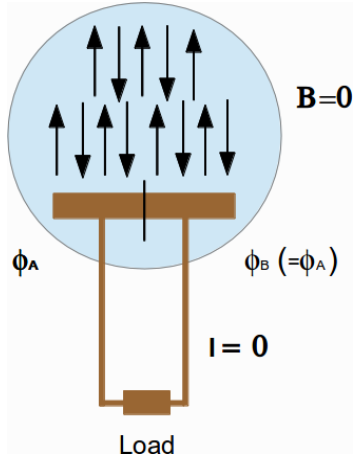}
\caption{Case of $\phi_A = \phi_B$ in Fu and Fu experiment. The arrows represent electrons in steady state equilibrium inside the bulb (space charge). In order to break the equilibrium, ${\bf B}$ is supposed to push the ``flying'' electrons from one side to the other, but this needs work and a static ${\bf B}$ cannot perform work (see Equation~(\ref{eq1})).}
\label{fig4}
\end{figure}

The static magnetic field can only modify the trajectory of each single electron in its movement back and forth between coatings and glass, but the overall equilibrium is preserved and the field cannot create a voltage drop and a current between the terminals of the sample (which would need external work). If we think of the steady state (back and forth) electron movements as two equal and opposite (compensating) currents, it would be like to assert that a static magnetic field is able to produce the Hall effect inside a conductor where two equal and opposite currents (\emph{i.e}.,~zero total current) flow.

In what follow, I outline two possible, mutually exclusive, scenarios to explain the experimental results described in the aforementioned papers. In the first scenario (I), the experimental results are not genuine and are the result of measurement flaws and/or environmental disturbances/interferences (natural and man-made electromagnetic ~waves, cosmic rays, or Thomson/Seebeck effects, \emph{etc}.). In the second scenario (II), the experimental results are genuine. In this case, I make reference to a more solid theoretical model in order to explain them.

\subsubsection{Scenario I}

The experimental results are not genuine and are the result of measurement errors and/or environmental disturbances/interferences. In this case, there is nothing really new to talk about. For people reading the papers~\cite{fu,pe}, it is actually difficult to evaluate possible procedural/measurements shortcomings. As a matter of fact, the experimental procedures followed by the authors of both experiments seem not to be completely ``watertight'' or, at the least, the authors have not been actually successful in describing them as ``watertight'' ones. For instance, I would have expected few simple checks in both cases, in order to exclude possible ``external'' sources of voltage/current. In the Fu and Fu experiment, I would have made a further test using a second bulb, identical to the sample one in all parts except the Ag--O--Cs coating of the electrodes. If a non-zero current could be detected also in this case, then the source would be probably in the interaction between measurement equipment (high sensitive electrometer) and the external magnetic field.

In the Perminov and Nikulov experiment they actually made some control measurements (zero magnetic field, inversion of the magnetic field) obtaining apparently consistent results, but the high temperature of the test environment ($\sim 100\,^{\circ}$C) and the high current used to generate the external magnetic field ($I_B\simeq 600$~A) leave me cautious about the definitive exclusion of possible interference/disturbance to the measurement equipment.

\subsubsection{Scenario II}

The experimental results are genuine. This second scenario is undoubtedly the most interesting and exciting one. Obviously, this poses an interesting theoretical challenge. I have shown before that the theoretical justification provided by the authors (the origin of the current/voltage drop is in the ``symmetry breaking'' by the external static magnetic field) is trivially flawed (see Equation~(\ref{eq1})). Here I suggest that the origin of the current/voltage drop may reside only in the work function difference of the electrodes. As a matter of fact, a quite solid and mature theoretical model of that already exists. As said before, I published a detailed physical/mathematical model of the thermo-charging process due to thermionic emission between electrodes with different work functions in~\cite{io4, io, io2, io2-1}.

Let me start saying something about the objections related to the voltage drop (and electric field) that allegedly would build up instantaneously between the free ends of two materials with different work functions just after being joined at one end, and that would prevent any thermionic emitted electron in the gap from spontaneously moving from low work function free electrode surface to high work function~one.

I have explicitly shown~\cite{io3} that no electric field, and thus no voltage drop, builds up between the surfaces at the free ends of two materials with different work functions when the materials are physically joined at one end. In~\cite{io3} I presented three arguments: the first two were more heuristic, the third one was more theoretical. Let me sketch here the last one, namely an explicit application of the {\em path-independence} law and/or Kirchhoff's loop rule. The physical principle at the basis of these two laws is the more fundamental law of conservation of energy. Conservation of energy demands that a test electronic charge $e$ conveyed around a closed path $\gamma$ in the device bulk of Figure~\ref{fig2}, through J-I (physical junction) and J-II (gap) at equilibrium, must undergo zero net work from {\em all} the forces present along the path. Mathematically, one must have,

\begin{equation}
\oint_{\gamma} d\textrm{W}_{ext}=0
\label{eq2}
\end{equation}

At equilibrium, the only two regions where forces are allowed to be non-null are the J-I and J-II regions. An electric field elsewhere in A and B (other than in the contact region) would generate a current, which contradicts the assumption of equilibrium. When the test charge $e$ crosses J-I, it is subject to the built-in electric field force $e\textrm{\bf E}_{bi}$ and to the diffusion force $\textrm{\bf F}_{diff}$. This last ``force'' is a thermally driven force and it is responsible for the establishment of the contact potential at J-I. We know that at equilibrium $e\textrm{\bf E}_{bi}=-\textrm{\bf F}_{diff}$ and that $\textrm{\bf F}_{diff}$ is different from zero and constantly present, otherwise $\textrm{\bf E}_{bi}$ would soon drop to zero, thus,

\begin{equation}
0=\oint_{\gamma} d\textrm{W}_{ext}=\int_{\textrm{\bf J-I}}
(e\textrm{\bf E}_{bi}+\textrm{\bf F}_{diff})\cdot d{\vec \gamma} +
\int_{\textrm{\bf J-II}}d\textrm{W}_{ext}=0+
\int_{\textrm{\bf J-II}}d\textrm{W}_{ext}
\label{eq3}
\end{equation}

In the J-II gap there are no diffusion forces ({{Remind that we ought not to consider thermionic emission ``diffusion forces'' in the gap since the alleged electric field the existence of which we want to dismiss is not that generated by thermionic emission; see beginning of Section~3.}}), since it is a vacuum gap, and eventually we have,

\begin{equation}
0=\int_{\textrm{\bf J-II}}d\textrm{W}_{ext}=\int_{\textrm{\bf J-II}}
e{\textrm{\bf E}_{\textrm{\bf J-II}}}\cdot d{\vec \gamma}= e|\textrm{\bf E}_{\textrm{\bf J-II}}|x_{g}
\quad\to\quad |\textrm{\bf E}_{\textrm{\bf J-II}}|=0
\label{eq4}
\end{equation}
where $x_{g}$ is the gap width.

Let me now analyse the Fu and Fu experiment under the light of that theory. I have said before that in their experiment one will hardly have exactly $\phi_A = \phi_B$. Thus, one has to assume $\phi_A < \phi_B$ or $\phi_A > \phi_B$.

Given that and according to the analysis done in~\cite{io4, io, io2, io2-1, io3}, even without the external static magnetic field, an equilibrium space charge eventually settles inside the bulb due to the thermionic emission. Under such an equilibrium, the electrode with lower work function ends up with an excess of positive charge with respect to the electrode with higher work function (which has an excess of negative charge). No built-in (contact potential generated) electric field between the free surfaces of the electrodes exists ({{See beginning of Section~3.}}) in advance that will prevent this final equilibrium. This means that, according to~\cite{io4, io, io2, io2-1, io3}, even without the external magnetic field a voltage drop settles between the two terminals of the bulb and a current flows when the terminals are shunted by a load. When the permanent magnet is moved close to the bulb, the magnetic field can only modify the geometry of the space charge distribution. This is as if the physical geometry of the electrodes (relative distance, relative orientation) were actually modified. In~\cite{io, io2, io2-1} I have already shown how the geometry of the electrodes, the relative distance and orientation influence the amount of current flow and consequently the voltage drop when the electrodes are shunted by the internal (high impedance) load of the electrometer. Please note that, contrary to what might appear, in doing that $\textrm{\bf B}$ does not perform work. We might also see the effect of the magnetic field as a manifestation of magnetoresistance.

This could explain the change from positive to negative of the current and voltage drop in the Fu and Fu experiment when the external magnetic field is reversed. Although in the paper by Fu and Fu it is not explicitly stated, I guess that they zeroed the electrometer connecting its terminals to the bulb in the absence of external magnetic field. But, according to what I have just said, this zero is not an actual zero. The reversing of the magnetic field does not actually reverse the flow of electrons from one electrode to the other, rather it changes the space charge distribution increasing or decreasing the one-way current. However, the zero of the measurement equipment being not the actual zero, such an increase or decrease above or below the relative zero is read as positive or negative by the electrometer.

%%%%%%%%%%%%%%%%%%%%%%%%%%%%%%%%%%%%%%%%%%%%%%%%%%%%%%%%%%%%

\section*{Acknowledgements}
\vspace{12pt}

This work has been partially supported by the National Institute for Astrophysics (INAF) under grant ``Bando Ricerca IAPS-2013''. Comments and suggestions by three anonymous reviewers are also~acknowledged.

%==========================================================
%==========================================================
% Back Matter (References and Notes)
%----------------------------------------------------------
% Style and layout of the references
\bibliographystyle{mdpi}
\makeatletter
\renewcommand\@biblabel[1]{#1. }
\makeatother
%----------------------------------------------------------
% Use the following option to include external BibTeX files:
%\bibliography{template}
%----------------------------------------------------------

\end{document}